\documentstyle[12pt]{article}
\begin{document}
\title{Relativity at Action or Gamma-Ray Bursts}
\author{Tsvi Piran  \\
The Racah Institute for Physics, \\
The Hebrew University, Jerusalem, Israel}
\maketitle

\begin{abstract}

Gamma ray Bursts (GRBs) - short 
bursts of few hundred keV $\gamma$-rays  - 
have fascinated astronomers since their
accidental discovery in the sixties.  GRBs  were  ignored by most
relativists who did not expect that they are associated with any
relativistic phenomenon. The recent observations of the BATSE detector 
on the Compton GRO satellite have  revolutionized out ideas on these
bursts and the picture that emerges shows that GRBs are the most relativistic
objects discovered  so far.
\end{abstract}

GRBs are short bursts of hundred keV range $\gamma$-rays  lasting from 
a few milliseconds to several hundred  seconds. 
GRBs were discovered accidentally by defense
satellites - the Vela satellites - that were launched to monitor the 
``outer space treaty" the 1962 non-proliferation treaty that forbade
nuclear explosions in space.
The discovery was kept secret
for a while and only in 1973 it was announced publicly \cite{Kle}.  
During the eighties a consensus formed that GRBs originate
on galactic disk neutron stars. In 1990 NASA launched Compton-GRO (Gamma-ray observatory)
satellite that  includes  BATSE a  GRB detector on board. 
BATSE which is more sensitive than any detector 
flown before detects on average one burst per day. BATSE 
was expected to find a  concentration of GRBs towards
the galactic plane and to prove the galactic neutron star model.
Instead BATSE discovered that the
GRB distribution has a perfect isotropy. Additionally BATSE detected a paucity
of weak bursts - the number of weak bursts, $N$,  did not increase with decreasing 
count  rate, $C$, like $C^{-3/2}$, as expected in a Euclidean space
\cite{Meegan92}. The simplest
explanation  is that GRBs are cosmological. 
Isotropy of cosmological
sources is obvious. As for 
the paucity of weak bursts here we encounter relativity for the first time. 
In a FRW universe  the volume
element does not increase like $r^3$ and the count rate from a given source
does not decrease like $r^{-2}$ (see e.g. \cite{weinberg}). 
If GRBs are detected  to $z\approx 1$ than these relativistic effects reduce 
significantly  the number of weak bursts in agreement with the 
observed $N(C)$ curve \cite{Pi92,CP95}.

Other cosmological effects should be observed as well. 
The spectrum of dim  (and hence distant) 
GRBs should  be red-shifted and the average duration of dim bursts
should be longer then the average duration of bright ones \cite{Pi92,Pac92}.
Lacking any clear spectral feature  (there are no spectral lines in GRBs)
it is difficult to observed the expected red-shift. However, time dilation of
dim bursts relative to bright ones have been recently reported \cite{Norris94}
confirming the theoretical  predictions and the cosmological origin of GRBs.

If GRBs are located at cosmological distances they release $\approx
10^{51}$ergs. Thus, they are the most (electromagnetically) luminous
objects known (only supernovae are more luminous, but their emission
is mostly neutrinos).  Already in the seventies Schmidt \cite{Schmidt}
pointed out that a compact source (as implied by the rapid 
variability) will be optically thick to pair creation
($\gamma\gamma \rightarrow e^+ e^-$) and consequently  it won't be
able to produce such a high luminosity  non-thermal radiation.  This
is the compactness problem.  

Relativistic effects can fool us and, when
ignored, lead to wrong  conclusions. This happened 30 years ago when
rapid variability implied ``impossible'' temperatures in
extra-galactic radio sources.  This puzzle was resolved when M. J.
Rees suggested ultra-relativistic expansion which has been confirmed
by VLBA measurements of super luminal jets.  This also happened in the present 
case. It has been argued on the
basis of the compactness problem that ``new physics" is unavoidable if GRBs are
cosmological. However, special relativity is all that is needed to resolve  this paradox
\cite{KroPie,RM,Pi94}. 
Consider a source of radiation that is moving towards an observer at rest with
a relativistic velocity characterized by a Lorentz factor,
$\gamma=1/\sqrt{1-v^2/c^2} \gg 1 $.  The size of the region 
from which the radiation is emitted should satisfy   $R_i <
\gamma^2 c \Delta T$  instead of the Newtonian estimate, $R_i < c \Delta T$. 
Additionally the photons with observed  energy $h \nu_{obs}$ have been blue shifted and their
energy at the source was $ h \nu_{obs}/\gamma$.  
The fraction of photons with sufficient energy to produce pairs
is smaller by a factor $\gamma^{-2\alpha}$ [where $\alpha$
is the spectral index]  than the observed fraction. 
These two effects reduce the optical depth by  $\gamma^{4+2 \alpha}$.
The system will become optically thin and the compactness problem
will be resolved if the emitting regions  move with an ultra-relativistic
velocity with $\gamma > 100$ ($v >0.99995c$). This is the largest observed
macroscopic velocity in the Universe - how can it be reached?

We do not expect macroscopic objects to roam in the Universe with velocities
approaching the speed of light. Instead the natural interpretation of this
conclusion is that this is an internal motion that is produced within 
the sources. It is the most (energetically as well as conceptually)
economical  to continue and suggest that the  kinetic energy of this
ultra-relativistic motion is also the source of the energy of the observed
GRBs - in other words that  GRBs are produces  during the slowing down
of a bulk motion of ultra-relativistic particles. This suggests a three stage process:
First,  an inner compact source  produces the energy. Second, 
this energy is transported
to a large distance not as $\gamma$-rays but  as the kinetic energy of an
ultra-relativistic particle flow  \cite{SP,Pac90,MR1}.  Third, the 
kinetic energy is converted to the observed radiation
only after it reaches a large enough distance where the emitting region
is optically thin.

One may wonder what produces the  ultra-relativistic particle flow.
The answer is quite simple. Suppose that a compact source with a radius $R_0$ 
emits radiation with a total energy $E$.
This radiation is optically thick and it produces what
has been called a ``fireball" - an optically thick sphere of radiation and 
electron-position plasma. The  fireball is initially  radiation dominated 
and it expands and accelerates \cite{Goo86,Pac86}. This phase resembles the 
early radiation dominated Universe. The radiation cools during the expansion
and its internal energy decreases until it is smaller than the rest mass energy
density of baryons that are present. The fireball reaches then 
a matter dominated stage in which the kinetic energy of baryons equals to the initial
energy of the fireball.
The baryons coast at a constant velocity with a Lorentz factor $\gamma = E /M c^2$
and their motion
is described well by a Milne Universe. From the point of view of an observer
at rest the baryons constitute a narrow shell with 
a thickness $R_0$ - the initial size of the fireball.  The kinetic energy
is transported outward where it has to be converted back to observed radiation.

It is worthwhile to consider a related well known phenomenon: supernova
explosion.  $10^{51}$ergs are deposited in a supernova into 
a stellar envelope of several solar masses, $M_{sn}$. One percent of the total
energy,
$\approx 10^{49}$ergs, is  observed  as the spectacular 
optical radiation of the supernova.
The
rest is the kinetic energy of the ejecta that moves at a velocity  
$u = \sqrt{2E /M_{sn}} \approx 10000$km/sec. 
When the ejecta  reaches a radius $R_{snr} = 10^{18}$cm
within which there is an equal mass of interstellar matter (ISM):
$(4 \pi/3)  R^3_{snr} \rho_{ISM} = M_{sn}$ it slows down
and  a supernova remnant (SNR) forms. The kinetic energy of the ejecta 
is then converted  to
non-thermal radiation over a period of $R_{snr} /u$ or tens of thousands of years. 

Imagine now that the same energy is deposited  into a much smaller mass
($M \approx 10^{-5} m_\odot$ or less). 
The ejecta will reach ultra-relativistic motion 
with $\gamma = E /Mc^2 > 100$. Now  effective
slowing down of the ejecta - that is slowing down from $\gamma$ to $\gamma /2$ -
takes place at  $(4 \pi/3)  R^3_{\gamma} \rho_{ISM} = M/\gamma = E / \gamma^2 c^2 $
(note that  relativistic effects  add a
factor of $\gamma$ in this relation).  Typical values of $R_\gamma$ are 
approximately $10^{15}$cm. The energy conversion takes place 
on a scale of $R_\gamma/c$ which is several days but it is observed
within $R_{\gamma} / \gamma^2 c$, 
which is sufficiently short if $\gamma > 100$. Like in SNR two shocks appear
in the interaction between the ejecta and the ISM: 
an outward going shock that propagates into the ISM and an inward going one 
that propagates into the ejecta. The 
shocks are ultra-relativistic and  the emitted radiation is much 
harder than in SNRs. The emitted photons are blue shifted (since the shocked material
moves relativistically relative to an observer at rest) and the observed 
photons are $\gamma$-rays in the hundred keV range.

So far we have seen cosmological effects and special relativistic effect 
- but this is only part of the  story. From the generic 
GRB model described above it appears that the actual source that produces
the energy and drives the whole process is well hidden and  GRB observations do not
reveal its nature. All that we know with certainty  
is that the sources should be  capable of producing $10^{51}$ergs within a 
short time and
that the process  takes place at a rate of approximately once per million
years per galaxy. The latter fact is obtained from estimates of the rates
of GRBs \cite{Pi92,CP95}. 
At present this circumstantial evidence is all that we know!

Among all astronomical objects binary neutron star mergers can be singled out
as candidates that satisfy both conditions \cite{Ei}. Binary neutron stars coalesce 
releasing $\approx 5 \times 10^{53}$ergs -  the binding energy
of a neutron star. Most of this energy escapes in the form of neutrinos or
gravitational radiation. Less than one percent of this amount  is sufficient to power a GRB. 
There are several models that suggest how such 
a fraction can be converted to electromagnetic radiation \cite{Ei,NPP}. The rate of neutron 
star mergers can be estimated  from binary pulsar observations.
Two independent estimates \cite{NPS,Phi} based on the  three known binary pulsars find
that the rate of binary neutron star mergers in our galaxy is one per million years.
The agreement with the observed rate of GRBs makes binary neutron
star mergers a prime candidate for sources of GRBs. In fact it is the only 
GRB model today that is based on an independently observed phenomenon that is
known to take place at a comparable rate.

Numerous interesting relativistic effects take place in binary neutron star mergers.
First, the merger is driven by gravitational radiation. The orbital decay due
to gravitational radiation emission was already observed in one binary pulsar,
PSR 1913+16 \cite{Taywei} and it manifests one of the greatest accomplishments
(both in terms of precision measurements and of theoretical predictions)
of general relativity so far. Second,  the combined mass of  
two neutron stars is most likely above the maximal mass of  a rapidly rotating
neutron star (which is uncertain due to uncertainty in neutron stars' equation
of state). Thus, the ultimate product of the merger is a rotating
black hole. If this model is correct then each GRB signals to us the formation
of a black hole! Other effects are still unknown or are just being explored now.
Complete modeling of binary neutron star mergers requires the most general gravitational
computer code: a dynamical three dimensional code that allows gravitational radiation 
emission as well as strong fields and black hole formation. 
Such codes do not exist yet but 
preliminary calculations
with simpler three dimensional 
codes suggest that an intriguing instability (which resembles the 
instability around marginally stable orbits in black holes) exists \cite{MW}.
This instability makes 
both neutron stars unstable  when they reach a certain distance from each other 
even before they collide. If this is true then the whole system may collapse
to a black hole  before a physical collision takes place. 
It is intriguing question whether GRB can form if such an instability
exists (note that if enough material is tidally torn from the neutron
stars before this catastrophe happens its accretion on the newly formed
black hole could power the GRB).

Searchers of gravitational radiation has already realized 
the importance of binary neutron star mergers. These  are  
the prime targets of the next generation gravitational radiation 
detectors  LIGO and VIRGO. 
If GRBs are produced in binary neutron 
star mergers then the accompanying GRB should 
increase the statistical significance of a gravitational radiation signal \cite{KP}. 
This would increase 
the sensitivity of a gravitational radiation detector by a factor of 2 and the  
detection  rate by half an order of magnitude. 
Clearly such a coincident detection will 
resolve the enigma that surrounded GRBs for almost three decades. 
This would be a nice small bonus to the achievement of directly establishing the
existence of gravitational radiation. It is amusing to note that such a joint
detection (in which the gravitational radiation from the orbital decay is expected
to precede the GRB by a fraction of a second) would immediately establish, using
the time of flight argument, a direct limit
on the graviton's mass. For example the arrival of a 100Hz gravitational radiation 
signal one second before a GRB signal will indicate that the graviton's  mass
is less than $10^{-21}$eV!

Quite unexpectedly and practically without notice  GRBs became 
the most relativistic objects known to us today: They demonstrate the existence of 
ultra-relativistic motion with velocities and 
Lorentz factors far larger than seen elsewhere 
in the Universe and they display 
time dilation and other cosmological effects. If GRBs 
are indeed produced in binary neutron star mergers (as seems most
likely today) than
they involve, gravitational radiation, strong dynamical gravitational 
fields and black hole formation. We may be seeing in GRBs the echoes of GR in full action.

This research was supported by a grant from the Israeli national science foundation.

{}
\end{document}